# Granular Segregation Mechanisms by Cyclic Shear

Zhifeng Li[1], Zhikun Zeng[1], Yi Xing[1], Jindong Li[1], Jie Zheng[1], Qinghao Mao[1], Jie Zhang[1,2], Meiying Hou[3] and Yujie Wang[1]*

[1]*School of Physics and Astronomy, Shanghai Jiao Tong University, 800 Dong Chuan Road, Shanghai 200240, China*
[2]*Institute of Natural Sciences, Shanghai Jiao Tong University, Shanghai 200240, China*
[3]*Institute of Physics, Chinese Academy of Sciences, Beijing 100190, China*

Granular mixtures with size difference can segregate upon shaking or shear. However, the quantitative study of this process remains difficult since it can be influenced by many mechanisms. Conflicting results on similar experimental systems are frequently obtained when the experimental conditions are not well controlled, which is mainly due to the fact that many mechanisms can be at work simultaneously. Moreover, it is often that macroscopic or empirical measures, which lack microscopic physical bases, are used to explain the experimental findings and therefore cannot provide an accurate and complete depiction of the overall process. Here, we carry out a detailed and systematic microscopic structure and dynamics study of a cyclically sheared granular system with rigorously controlled experimental conditions. We find that both convection and arching effect play significant roles on the segregation process in our system and we can identify quantitatively their respective contributions.

Granular mixtures consisting of different size particles tend to segregate under shaking or shear, which has important implications for many industrial processes when either mixing or segregation is desired[1]. It is also relevant to geophysical processes like landslides, avalanches and block size distributions on asteroids[2,3]. Granular segregation can be induced by size, density, friction, inelasticity while size remains the dominant factor[4-9]. Owing to its ubiquitous existence and importance, it is crucial to gain a fundamental understanding of segregation mechanisms based on systematic experimental investigations. Continuous segregation models can subsequently be developed based on them[10-12]. However, the experimental investigations of granular segregation are highly nontrivial. This is due to the facts: (i) complex dynamics associated with shaking or shear cannot be easily visualized in three dimension (3D). Early investigations used empirical measures like particle segregation fraction or the time needed for particle to rise to surface[13]. These approaches were done at a cost without a deep understanding of the microscopic mechanisms[6]. X-ray tomography, magnetic resonance imaging, and refractive index matching method have provided some great insights in 3D segregation process[14,15]. However, more work is needed to combine microscopic structure and dynamics information to systematically investigate the underlying mechanisms. Until recently, most experimental investigations are limited to two-dimensional systems; (ii) many factors and mechanisms can contribute simultaneously to the segregation process. The most well-known factor is particle size, but density, the presence of interstitial fluid, or friction can also influence the segregation process[4,7,16-18]. Even when only size disparity exists, mechanisms including arching effect[19,20], convection[21], thermal gradient[22] have all been demonstrated to play important roles in the segregation process. By nature, any gradient in the system can lead to segregation and all these mechanisms can induce gradients in the system in certain way[23]. It is therefore vital to have controlled experiments which can separate their individual contributions.

In the present study, we employ X-ray tomography to investigate granular segregation in a quasi-statically sheared granular medium. When the system is under quasi-static shear and therefore particle velocity or kinetic energy is irrelevant, mechanisms like thermal diffusion[24,25] or kinetic mechanisms[9,22] can be excluded. This is in contrast to agitation by shaking in which gradients in both particle density and velocity exist. By systematically varying the tracer size, we obtain particle-level structure and dynamics. We find that convection and geometric arching mechanism are the two major mechanisms which contribute to segregation in the dense packing and dilute tracer limit.

**Results**

In our system, tracer particles segregate from background particles upon cyclic shear and reach steady state distributions when extensive shear cycles are applied. This segregation process is a combining effect of several segregation mechanisms working at different timescales ranging from single to thousands of cycles. It is experimentally very costly to do tomography scans covering the full timescale. To improve experimental efficiency, we follow three specific experimental protocols to address specific timescale properties of the system and corresponding segregation mechanisms (more details can be found in Methods section), which can significantly reduce the scan number needed. In the following, we show in detail how different mechanisms contribute together for the segregation process of our system.

**Convection induced global segregation.** One of the major mechanism proposed for granular segregation is convection[21]. By tracking the trajectories of the background particles using protocol B, we can obtain the global convection pattern as shown in Fig. 1c. The convection mainly consists of particles rising in the center and moving down in thin sheets at two shear walls and in funnel-like shapes at four corners. The cross-section of the downward convection shrinks continuously as depth increases. Fig. 1b shows the typical height (*H*) trajectories for $D = 12$ mm tracer particles as a function of shear cycle number *t* as obtained by protocol A. From the tracers' trajectories we can compile the tracer flow field as shown in Fig. 1d. It is clear that tracers can move both up and down by following loop-like trajectories. Overall, the tracer flow field is rather similar to the background convection field. It is therefore obvious that tracers flow with the convection. However, there exist subtle differences between two flow fields as big tracers cannot penetrate as deep down in the downward convection as the background particles. This is due to the fact that big tracers are unable to follow the downward convection flow as its cross-section shrinks with depth. The tracers are therefore carried up again by the upward convection at certain depth to form smaller convection rolls. Additionally, the tracer convection rolls will be mainly confined to four corners instead of traversing the whole volume. This phenomenon is more dramatic as we increase the tracer size. Fig. 2a-e show the snapshots of different size tracers' positions in the shear chamber when they have reached their respective global steady-state distributions. It is clear that tracers are now mainly trapped within their individual convection rolls, with smaller tracers penetrating deeper while bigger tracers remain at the surface, since their sheer size will prevent them from entering the downward convection roll. Fig. 2f show the probability distribution functions of different size tracer particles as a function of depth ($H_0$-*H*, $H_0$ is the surface height) at steady states. It clearly demonstrates that convection itself can lead to particle segregation. However, it is also clear that simple distributions will bring little insight on the mechanism of particle segregation.

**Local segregation.** It is interesting to see whether convection is the only mechanism for segregation in our system. Knight *et al.* observed that particles with different sizes will all rise with the same speed[21]. Vanel *et al.*[26], on the other hand, suggested that the tracer rises faster when the size is bigger. As shown in Fig. 3a, we monitor the heights of the center of mass (COM) for all tracer particles as a function of shear cycle number $t$. The heights of COM for different size tracers will reach steady state after about several hundred shear cycles. However, we can also notice that the COMs of large tracers rise faster than smaller ones at the beginning stage. The corresponding speeds $v_{COM}$ (in units of $d$/cycle) are shown in Fig. 3b. This result implies that larger tracers could move up faster than smaller ones. One thing worth noting is that we can even observe a much slower segregation happening within the background particles as the height of COM of 5 mm particles decreases slowly over time. Since COM is a global average including tracers moving downward, the information it carries could be ambiguous. Therefore, we analyze the relative motion between the tracer and its neighboring background particles. The analysis is carried out only during the period when the height of COM is still increasing and only for tracers that are moving upward. Using protocol B, we obtain the height trajectories of sixteen $D=12$ mm (Fig. 3c) and four $D=24$ mm (Fig. 3d) tracers as a function of shear cycle number $t$ and their corresponding relative height trajectories after subtraction of the average vertical displacements of their neighboring background particles within a distance of $2d$. The results are shown in the upper and lower insets of Fig. 3c, d respectively. The main panels of Fig. 3c, d show the average behavior of all tracers. It is found that even after subtraction, the tracers still have some net upward speeds as compared to their neighbors. The speed is larger for $D=24$ mm tracers than that of $D=12$ mm ones. It is clear that this relative speed comes from a different origin than convection. We denote $v$ as the tracer's vertical speed and $\langle v_{nei} \rangle$ as the average vertical speed of its neighbors within a distance of 2$d$. The normalized speeds of different size tracers $v/\langle v_{nei} \rangle$ are plotted in Fig. 3b. It is in general consistent with the COM behaviors. Since we are in quasi-static shear regime where particle kinetic energy is clearly irrelevant[24], and the relative displacements happen with the tracers' immediate neighbors, we have to find some local cause for this relative motion.

*Archimedean force.* We first exclude the possibility of Archimedean buoyancy force. Specifically, we calculate the local volume fraction $\phi = v_g / v_{voro}$ of each particle based on radical Voronoi tessellation, where $v_g$ and $v_{voro}$ are its respective volume and Voronoi cell volume[27]. We find that tracers possess much larger $\phi$ than those of the background particles, *e.g.*, $\phi = 0.88$ for $D = 12$ mm tracers and $\phi = 0.61$ for their neighbors. This observation is a direct proof of the irrelevancy of the Archimedean force since its effect in principle should lead the tracers to sink relative to their neighbors. This is contrary to the case in dilute system limit, where an effective buoyancy force might play a major role based on kinetic theory calculations[28].

*Local gradient and arching effect.* We further search for certain gradients in structure around the tracer which can lead to local segregation. As shown in Fig. 4b, we calculate the average local volume fraction $\phi$ for particles within 4$d$ of $D = 8, 12, 24$ mm tracers. The particles in the lower half have a lower $\phi$ than that of the upper half (*e.g.*, for $D = 12$ mm, the lower half $\phi = 0.593$ and the upper half $\phi = 0.604$ ), and the difference increases as the tracer size increases. This up-down asymmetry and its trend are also apparent in the contact distribution on the tracers. As shown in Fig. 4a, the average contact number $Z$ of an $D=12$ mm tracer is about 15. However, it is not evenly distributed in the upper and lower half hemispheres. The average contact number $Z$ of upper hemisphere is 8.41, and that of the lower hemisphere is 6.56. This clear uneven distribution of $\phi$ and $Z$ reminds us of the arch segregation mechanism[20]: owing to the presence of gravity and friction, granular particles can form arches or bridge structures which are collective structures where neighboring grains rely on each other for mutual mechanical support. The bridges are normally dome-like which will induce an up-down density asymmetry by shielding cavities or less dense regimes underneath them[29,30]. When cyclic shear is applied, the big tracers are supported by the arch structures while small background particles percolate into the voids or lower density regimes underneath them. This will result in a net upward displacement of the tracers versus their neighbors after one shear cycle. The details of this local segregation process can be elucidated by monitoring the dynamics and structure evolution of one 30mm tracer and its neighboring particles within one shear cycle by protocol C, and explained by the combined effect of bridge/arch structure (see Methods section for identification method of arch or bridge structures), density asymmetry, and particle local flow dynamics. As shown in Fig. 5b, during the first 1/4 shear cycle when the system is sheared rightward, the system is stretched in AC direction and compressed along BD direction. Mechanically rigid arch/bridge structures are formed preferentially along BD direction which can prevent the tracer from sinking. While the relative volume fraction $\phi_{rela}$, in the lower right corner along AC direction decreases gradually to form a mechanically unstable regime (Fig. 5a). $\phi_{rela}$ is defined as $\phi_{rela} = \phi - \langle\phi\rangle$ to remove the influence of global shear dilation, where $\langle\phi\rangle$ is the average volume fraction in bulk region (2$d$ away from the boundary of the shear box). Background particles tend to be fluidized to fill this corner, which generates their downward displacements relative to the big tracers along AC direction. In the second 1/4 shear cycle, the compression direction switches to AC direction and along which new bridge structures are gradually formed. The already filled-in particles push the tracer along upward AC direction since they are stuck in bridge structures and are difficult to return to their original positions before fluidization. Therefore, after the first half shear cycle, the tracer acquires a net displacement along upward AC direction relative to its neighbors. During the second half shear cycle, the behavior of the tracer and its neighbors is analogous which induces a net displacement of the tracer along upward BD direction. Consequently, this void-filling mechanism at the two lower corners (Fig. 5c) will lead to the vertical zigzag upward displacements of the tracer particle upon shear.

Although this mechanism works for all tracers, we find that larger tracers can sustain stronger arch structures than smaller ones. This is owning to the fact that background particles are mostly involved in simple linear bridge structures whose backbones do not have loops or branches. The large tracer, however, will turn the bridge structures containing it into a complex one since it can simultaneously mechanically support or be supported by many of its contact neighbors and therefore be part of many bridges. This will substantially extend the spatial extension and size of the bridge structure containing it as compared to the ones formed by background particles only. In fact, larger the tracer, more dramatic the effect is. Therefore, larger tracer particles will induce more significant up-down density asymmetry around them as compared to smaller ones, as shown in Fig. 4b. This could lead to bigger fluidization of neighboring particles which explains why larger tracers move faster than smaller ones.

Interestingly, contrary to the geometric model[19,20] originally proposed, in our system, there does not exist a size ratio threshold, when below this threshold, the local segregation ceases to exist[19,20]. Nevertheless, as shown in Fig. 3b, the local segregation speed increases as the tracer size increases, and it saturates around $D=18$ mm which is roughly consistent with threshold size ratio 2.8 predicted[19]. This indicates a transition to a limiting behavior where the background particles can be considered as a continuous medium. However, the existence of local segregation below this threshold points to important collective effects which are crucial for the formation of arches and might not be taken into account properly in previous simulations[19,20], as suggested by ref. [31].

*Global density gradient and depletion interactions*. We note that another possibility for segregation is that there exists global density gradient in the system. As shown in Fig. 4c, the gradient is indeed present. However, it is too small at the particle length scale to have a dominant effect over the arching effect. Until now, we have implicitly assumed that our system is in the dilute tracer limit. However, when tracer number density is high, potential depletion interactions among them can occur[32] and they can even form a rigid matrix for the small background particles to sieve through[11]. To investigate whether there exist correlations among tracers, we also reduce the tracer number from 200 to 20 in the $D=12$ mm case in protocol A. We do not observe any significant differences of two experiments. We also calculate the pair correlation function g(r) of the tracers, as shown in Fig. 4d, and find it is hard sphere like, thus excluding the existence of significant attractive depletion interactions. This can also be justified by the fact that in most of our experiments the tracer particles are spaced on average by 4d at which distance g(r) is already featureless.

**Friction effect.** Although we do not systematically investigate the effect of friction, we notice that friction plays an important role in our experiment[17]. When the experiment is performed over very long periods, the friction coefficient $\mu$ of particles reduce from 0.434 to 0.409 (by gluing three particles to make a sledge, $\mu$ is measured by the start sliding angle of the sledge on a tilted smooth plastic ABS plate, the same material as the background particles), and the convection speed decreases from 0.08$d$/cycle to 0.0033$d$/cycle, which significantly slows down the segregation rate.

**Discussion**

To summarize, we study the segregation process inside a bi-disperse granular system under quasi-static cyclic shear and find that convection and arching effect are the two major mechanisms for segregation in our system. This result is significant in the sense that it is obtained under rigorously controlled experimental conditions and provides quantitative microscopic information of the granular segregation process. In previous experimental studies, since any type of vertical gradient in granular system can cause segregation, it is often the case that contributing factors like size disparity, density, friction, shape, interstitial fluid, restitution coefficient and energy input method are at work simultaneously, which leads to rich and seemingly conflicting results when experimental conditions are not well controlled. Moreover, these studies normally only employ some macroscopic or empirical measures to characterize the segregation process, which lack microscopic physical bases, and therefore cannot provide an accurate and complete depiction of the overall process. These aforementioned issues have made the field progress at a slow pace. In the current work, we carry out a detailed and systematic microscopic structure and dynamics study of a quasi-statically sheared granular system with rigorously controlled experimental conditions, provide the relevant microscopic mechanism for segregation and clearly separate their respective contributions, which should pave the way for future quantitative study of granular segregation process.

*Acknowledgements*: *We would like to acknowledge useful discussions with Walter Kob. The work is supported by the National Natural Science Foundation of China (No. 11974240, No. 11675110, No. U1738120), Shanghai Science and Technology Commission (No. 19XD1402100).*
*Email*:*yujiewang@sjtu.edu.cn*

**Methods**

**Experimental sample and set-up.** The schematic design of the experimental setup is shown in Fig. 1a. The rectangular shear box is made of plexiglass plates and has a size of $120(L) \times 120(W) \times 140(H)$mm$^3$. The front and back plates are permanently fixed on the apparatus base. Two side plates and one bottom plate form a deformable *U*-shaped structure, where the upper edges of the two side plates are bolted on vertical slots on the front and back plates and the lower edges are connected with the bottom plate through hinges. When step motor drives the bottom plate to translate horizontally, the *U*-shaped structure will deform and generate shear on the system.

The background particles consist of 5 mm and 6 mm diameter ABS ($\rho = 1.0$ g cm$^{-3}$)

beads with 7,000 of each. The particles fill the chamber to a height of $H = 13.5$ cm with a free upper boundary and the particles are well-mixed to prevent crystallization. We denote by $d = 6$ mm the size of the larger background particle. The tracers are of the same material and surface properties as the background particles with diameter of $D$ = 8, 10, 12, 14, 16, 18, 24, 30 mm respectively.

Cyclic shear is applied with a strain amplitude of $\gamma = 0.33$ and a shear rate of $\dot{\gamma} = 0.33/s$. The corresponding inertial number is $I = \dot{\gamma}d/\sqrt{P/\rho} = 1.8 \times 10^{-3}$, therefore ensuring the shear quasi-static. We estimate the pressure by $P = \phi\rho gH$, with the volume fraction $\phi=0.6$. Upon cyclic shear, the bottom plate first moves in one direction and tilts the shear box into a parallelepiped until a target shear strain is reached. Then the shear direction is reversed, tilting the shear box into a symmetric parallelepiped shape with the same target shear strain in the opposite direction. Finally, the shear is reversed again and the shear box is returned to its original geometry to complete one full shear cycle.

The three-dimensional structural evolution of the granular particles inside the shear box upon cyclic shear is acquired by a CT scanner (UEG Medical Imaging Equip. Co. Ltd.). The spatial resolution of the CT scan is 0.2 mm. Following the similar image processing techniques as described in previous studies[33], coordinates and size of each particle can be obtained with an error less than $3\times10^{-3}d$. Once we acquire the coordinates and size of each particle, we can obtain their trajectories through a tracking algorithm by analyzing consecutive CT scans: a particle in the first scan is considered to be the same particle in the second scan which has the closest spatial location to it. This tracking algorithm works only if the typical displacements of particles are much less than the average inter-particle distance. In practice, in order to track the background particles, the particle displacements have to be less than half of the background particle diameter.

**Experimental protocol.** Different experimental protocols are used to address different timescale properties of the system as well as different segregation mechanisms. Specifically, the following three experimental protocols are adopted.

*Protocol A.* We use protocol A to track the tracers' positions in the convection rolls on the timescale of tens to thousands of shear cycles. It is necessary to follow the tracers' trajectories for a long period of time since we need the tracers to reach a global steady-state distribution for us to investigate convection-induced segregation effect.

We deposit tracers with one specific size into the bi-disperse background particles when we first prepare the system. The tracers are carefully deposited into two layers at specific depths, which are about 2cm away from the bottom and the surface respectively. To maintain a roughly same total volume of the system for different tracers, we fix the number of background particles and change the number of tracer particles accordingly. Specifically, $N_{8mm}$=642, $N_{10mm}$=294, $N_{12mm}$=200, $N_{14mm}$=134, $N_{16mm}$=74, $N_{18mm}$=62, and $N_{24mm}$=24 tracer particles are used. One CT scan is taken for every 30 consecutive shear cycles and a total of 100 CT scans are taken for each tracer size. The tracers' trajectories can be tracked since they are spatially separated from each other, while those of the background particles cannot.

*Protocol B.* The local segregation mechanism happens at the time scale of single shear cycle level. We use protocol B to track the trajectories of both the tracers and their neighboring background particles for every shear cycle.

Similar to protocol A, tracers with specific size are used. For different experiments, $N_{8mm}$=135, $N_{10mm}$=70, $N_{12mm}$=40, $N_{14mm}$=25, $N_{16mm}$=17, $N_{18mm}$=12, $N_{20mm}$=9, $N_{24mm}$=5 and $N_{30mm}$=3 tracers are used and they are deposited randomly inside background particles. After deposition, we take one CT scan after each shear cycle and a total of 200 CT scans are taken for each tracer size.

In this protocol, in the bulk region where particles are at least $2d$ away from the boundary of the shear box, the displacements of all particles after one shear cycle are less than $1/2d$ and therefore all particles' trajectories can be tracked. Close to shear box boundary, since particle convection speeds are significantly larger than those in the bulk region, their trajectories cannot be tracked.

*Protocol C.* Using Protocol B, we can establish a strong correlation between the arch-induced up-down volume fraction $\phi$ asymmetry around tracers and the local segregation speed, which suggests the importance of void-filling mechanism for local segregation. However, the specific mechanism that void-filling works in 3D to induce segregation needs to be clarified. In protocol C, single step shear experiment of one shear cycle is carried out in which we analyze the evolution of volume fraction, bridge structure and local flow dynamics around the tracers within one shear cycle to demonstrate how the existence of arches can lead to local segregation. To monitor the structure evolution and dynamics within one shear cycle, we divide one shear cycle into 160 shear steps and take a CT scan after each shear step. We deposit one 30mm tracer particle initially in the middle of the *xy* plane at the bottom of the shear box. Then we apply a 50 cycles' shear to prepare the system and also through which to move the tracer particle to a height of around $10d$ from the bottom. Subsequently, we carry out single step shear experiment for 30 consecutive shear cycles.

In this protocol, owing to the small shear step, the displacements of all particles are less than $1/3d$ and all particles can be tracked.

**Identification of the bridge structure**. We follow standard procedures to identify bridge structures in our system[34]. To identify bridges, the contacts between particles have to be determined first since bridges are by definition collective structures where neighboring grains mechanically support each other through contacts. We follow standard procedures to determine the particle contacts in our system by complementary error function fitting for our CT acquired packing structures[35]. Once the contact network is identified, the next critical step is the identification of the force-bearing neighbors among all contact ones, from which the bridges can be determined: a mechanically stable particle under gravity is generally considered to be supported by a base of three contact neighbors, with the requirement that the projection of the particle's center of mass falls within the triangle formed by three base particles. As there exist many possible combinations of three particles satisfying the support-base requirements above, we identify the effective base by using the standard "lowest center of mass (LCOM)" method which chooses the support base as the one possessing the lowest average centroid among all possible bases[34]. The particles in the base that are mutually supportive for each other form the bridge structure. Our identification of arch or force-bearing structures using the bridge concept is consistent with our experimental observations of local segregation process, which justifies above analysis.

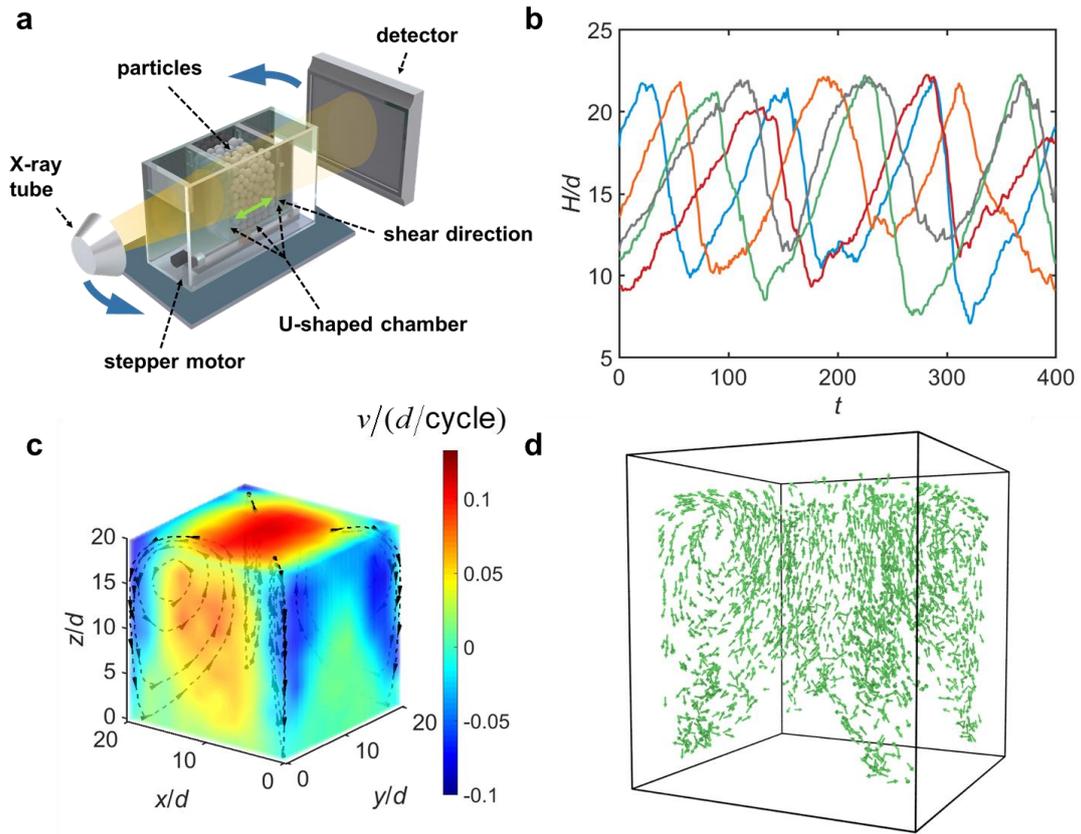

**Figure 1 | Experimental set-up and convection pattern. a,** The experimental setup. **b,** Tracers' height trajectories as a function of shear cycle number $t$. **c,** The flow pattern of convection. **d,** The flow pattern of tracer particles ($D=12$ mm) at steady state.

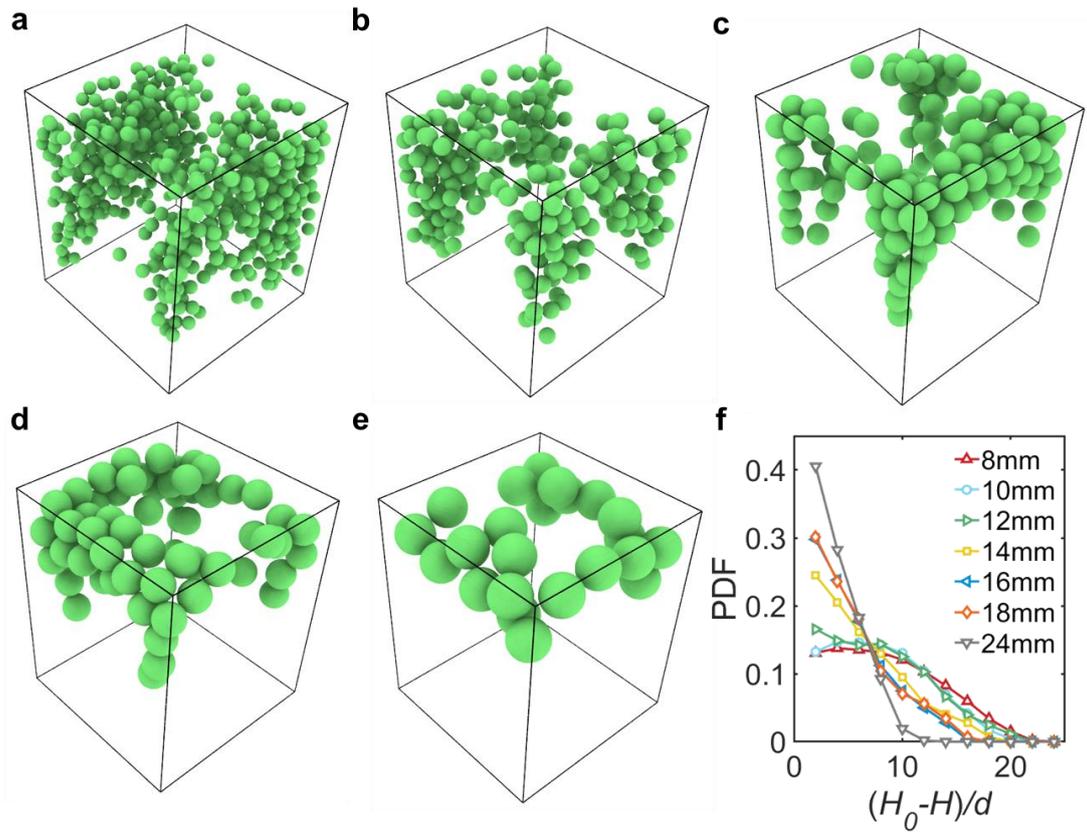

**Figure 2 | Snapshots of different tracer particles at their respective steady states.** The snapshots of particle positions of **a,** 8 mm, **b,** 10 mm, **c,** 12 mm, **d,** 16 mm, **e,** 24 mm tracers at steady states. **f,** Probability distribution functions of tracer particles as a function of depth at steady states.

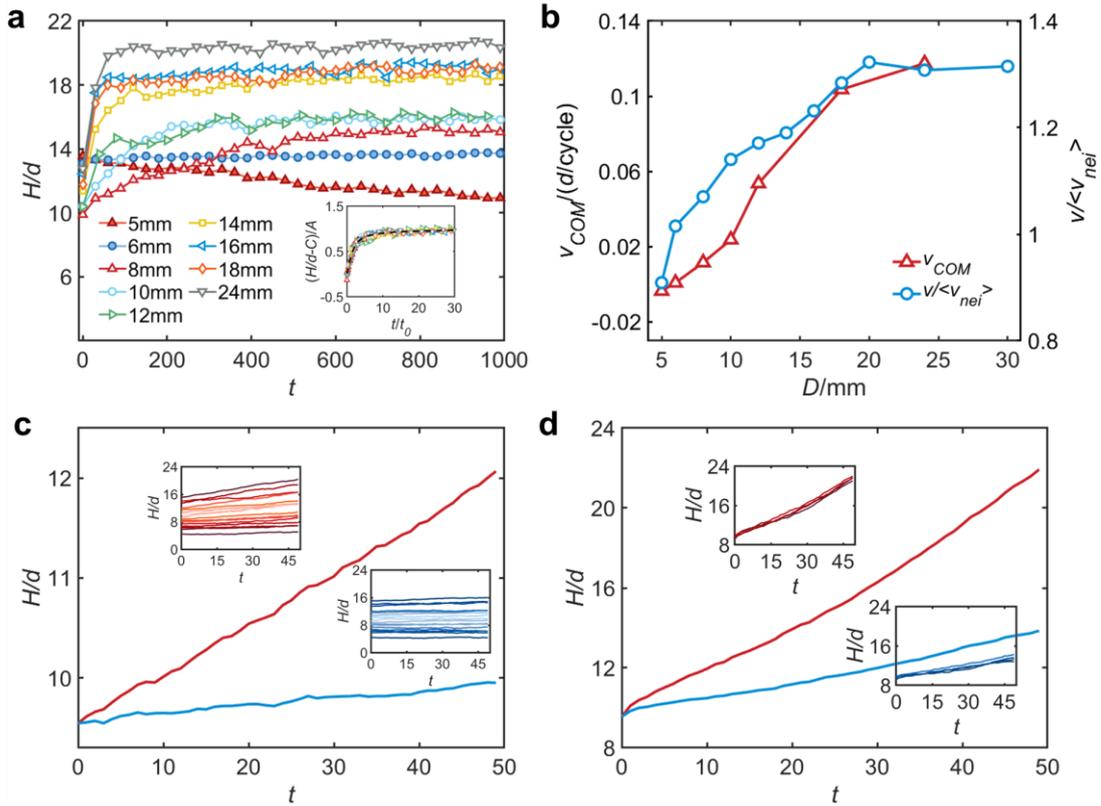

**Figure 3 | Height trajectories and segregation speeds of tracer particles. a,** The heights of the COMs for different size particles as a function of shear cycle number $t$. Inset: All curves can be fitted by $H/d = A\exp(-t_0/t) + C$, where $t_0$ is the intrinsic time scale of different size particles to reach the steady states. All curves can collapse after rescaling with $t_0$. **b,** The average speeds of the COMs $v_{COM}$ (red, left axis) and the average normalized speeds $v/\langle v_{nei}\rangle$ (blue, right axis) for different size particles. **c** and **d,** The height trajectories of sixteen $D = 12$ mm and four $D = 24$ mm tracer particles as a function of shear cycle number $t$. The upper and lower insets show the absolute (red) and relative (blue) height trajectories of the tracer particles before and after subtraction of the neighboring particles' vertical displacements.

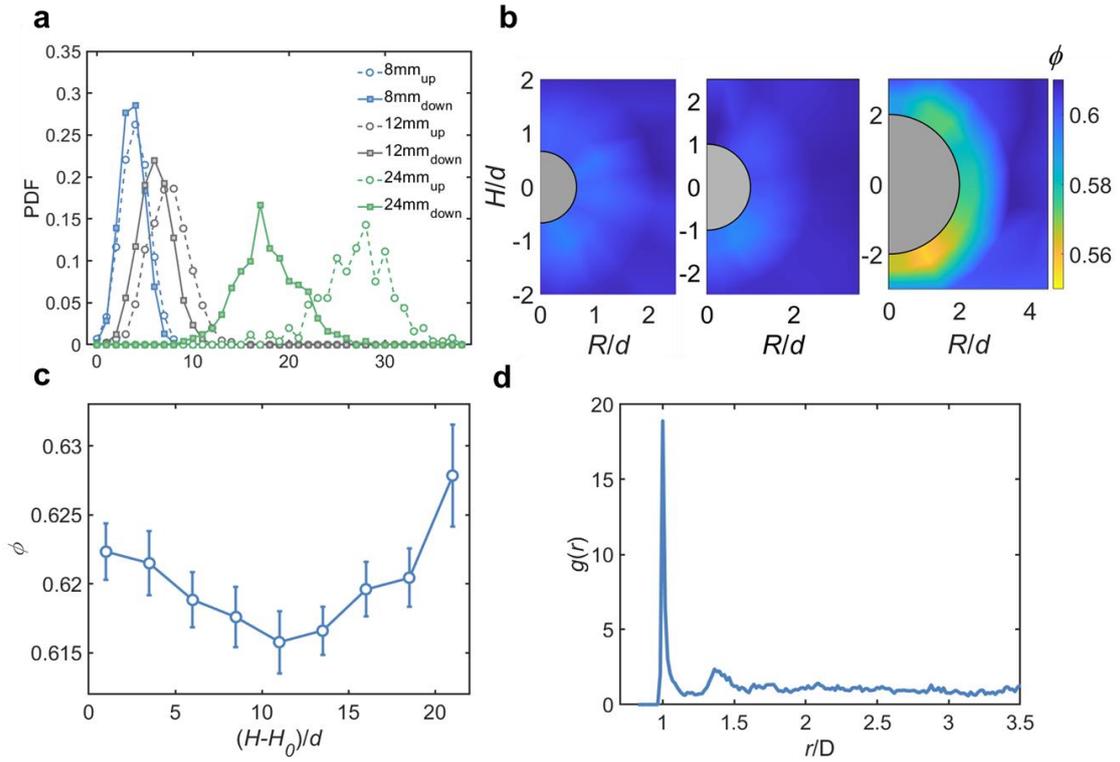

**Figure 4 | Up-down asymmetry of *Z* and *ϕ* around the tracer particles, global density gradient and the pair correlation function of tracer particles. a,** The probability distribution functions of contact number in the upper and lower hemispheres for 8, 12, 24 mm tracers. **b,** The average volume fraction distribution of background particles within 4*d* distance to 8, 12, 24 mm tracers which show the up-down asymmetry. **(c)** The average volume fraction *ϕ* of the system as a function of depth. **d,** Pair correlation function between 200 12 mm tracers.

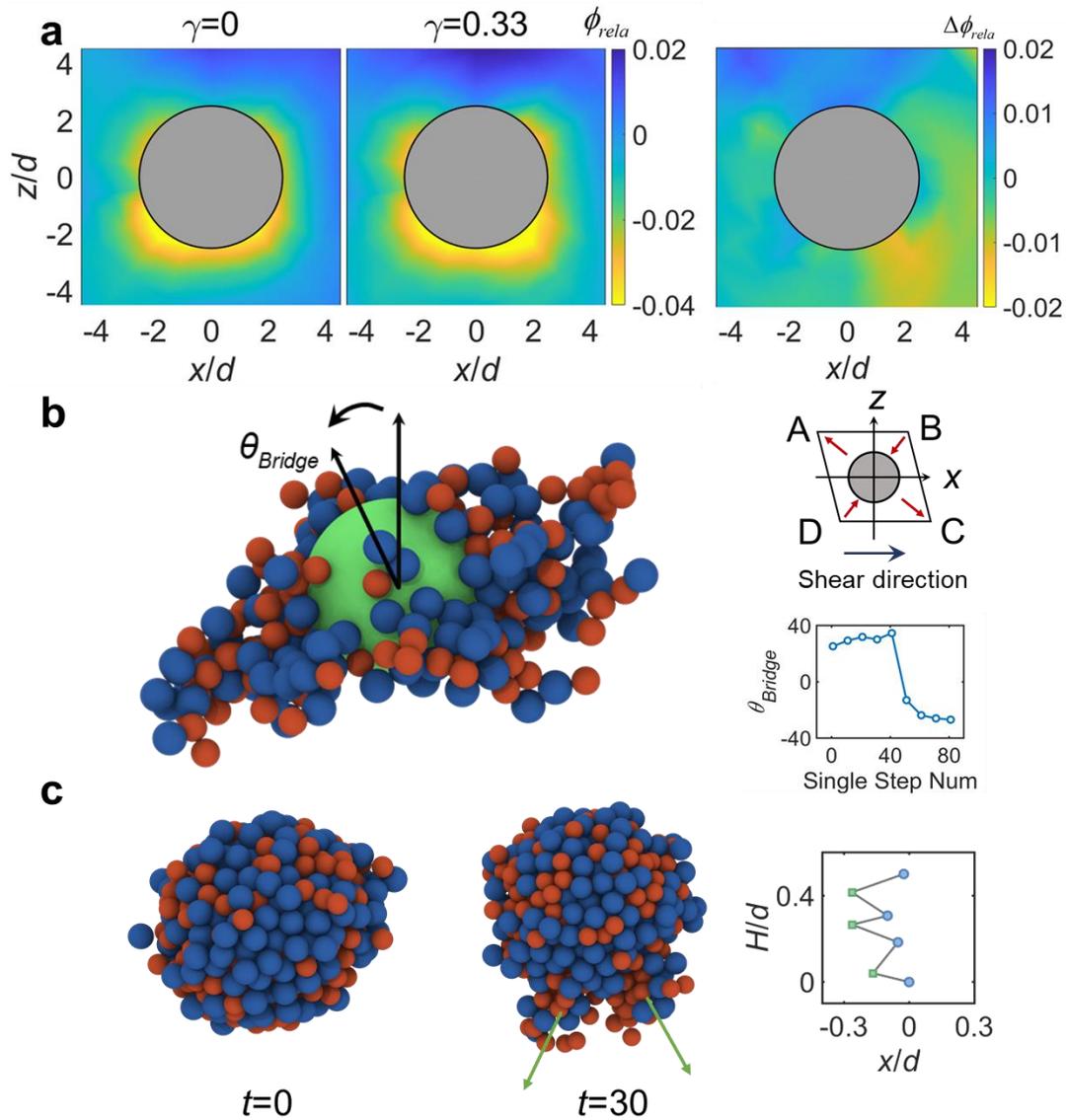

**Figure 5 | Arching effect and fluidization around tracer particles within one cycle.**

**a,** The relative volume fraction $\phi_{rela}$ around the tracer at $\gamma=0$ (left) and $\gamma=0.33$ (middle); Right: The variation of $\phi_{rela}$ around the tracer $\Delta\phi_{rela} = \phi_{rela,\,\gamma=0.33} - \phi_{rela,\,\gamma=0}$ between $\gamma=0.33$ and $\gamma=0$ during the first 1/4 shear cycle. **b,** Left: The complex bridge structure containing the tracer when the system is sheared rightward. The orientation of the complex bridge structure is defined as the orientation of the principal axis of its inertia tensor with the maximum eigenvalue. $\theta_{Bridge}$ is the angle between the orientation of the bridge structure and z-axis on the xz plane. Upper right: The schematic diagram of rightward shear. Lower right: The evolution of $\theta_{Bridge}$ during the first half shear cycle. When shear is reversed, $\theta_{Bridge}$ changes its direction from BD ($\theta_{Bridge}>0$) to AC ($\theta_{Bridge}<0$). **c,** Left: Background particles around the tracer at the initial state (5 and 6 mm particles are colored by red and blue); Middle: The same particles as the left panel after 30 shear cycles. The green arrows represent the preferred directions that particles flow relative to the big tracer; Right: The trajectory of the tracer particle acquired at the interval of half shear cycle. The zigzag shape demonstrates the asymmetric arch effect due to shear.